\documentclass[%
 aps,pra,reprint,
 superscriptaddress, 
 showpacs,showkeys,
 noeprint,nopreprintnumbers,
 amsmath,amssymb,
 longbibliography,
 nofootinbib,
 floatfix
]{revtex4-1}
\usepackage[utf8]{inputenc}
\usepackage[pdftex]{graphicx}
\usepackage[usenames, dvipsnames]{xcolor}

\usepackage{braket} 
\usepackage[normalem]{ulem} 
\usepackage{dcolumn}
\usepackage{bm}
\usepackage{bbold} 
\usepackage{hyperref} 
\usepackage{comment}
\usepackage{bibunits}
\usepackage{ dsfont }
\usepackage{ mathrsfs }
\usepackage{cancel}
\usepackage{xparse}
\usepackage{physics}
\usepackage{soul}
\usepackage{bibunits}
\defaultbibliographystyle{apsrev4-1}
\defaultbibliography{cavity-biblio}

\graphicspath{ {./IMG/} {./} }
\newcommand{\ksig}{\ensuremath{{\bf k},\sigma}}

\newcommand{\fir}[1]{Fig.~\ref{#1}}

\newcommand{\eqr}[1]{Eq.~(\ref{#1})}

\begin{document}

\def\mytitle{Photoinduced electron pairing in a driven cavity
}
\title{\mytitle}

\author{Hongmin Gao}
 \email{hongmin.gao@physics.ox.ac.uk}
 \affiliation{Clarendon Laboratory, University of Oxford, Parks Road, Oxford OX1 3PU, United Kingdom}

\author{Frank Schlawin}
 \email{frank.schlawin@physics.ox.ac.uk}
 \affiliation{Clarendon Laboratory, University of Oxford, Parks Road, Oxford OX1 3PU, United Kingdom}

\author{Michele Buzzi}
  \email{michele.buzzi@mpsd.mpg.de}
  \affiliation{Max Planck Institute for the Structure and Dynamics of Matter, Hamburg, Germany}

\author{Andrea Cavalleri}
 \email{andrea.cavalleri@mpsd.mpg.de}
 \affiliation{Clarendon Laboratory, University of Oxford, Parks Road, Oxford OX1 3PU, United Kingdom}
 \affiliation{Max Planck Institute for the Structure and Dynamics of Matter, Hamburg, Germany}

\author{Dieter Jaksch}
 \email{dieter.jaksch@physics.ox.ac.uk}
 \affiliation{Clarendon Laboratory, University of Oxford, Parks Road, Oxford OX1 3PU, United Kingdom}

\date{\today}

\begin{abstract}
We demonstrate how virtual scattering of laser photons inside a cavity via two-photon processes can induce controllable long-range electron interactions in two-dimensional materials. 
We show that laser light that is red(blue)-detuned from the cavity yields attractive(repulsive) interactions, whose strength is proportional to the laser intensity. Furthermore, we find that the interactions are not screened effectively except at very low frequencies. For realistic cavity parameters, laser-induced heating of the electrons by inelastic photon scattering is suppressed and coherent electron interactions dominate. When the interactions are attractive, they cause an instability in the Cooper channel at a temperature proportional to the square root of the driving intensity. 
Our results provide a novel route for engineering electron interactions in a wide range of two-dimensional materials including AB-stacked bilayer graphene and the conducting interface between $\text{LaAlO}_3$ and $\text{SrTiO}_3$. 
\end{abstract}

\maketitle

\begin{bibunit}

\nocite{apsrev41Control}

\paragraph*{Introduction.}
Engineering material properties on demand is widely considered as one of the central goals of modern condensed matter physics \cite{Basov17}. 
This can be achieved by static means, for instance by exfoliating materials into single atomic layers and combining them into van der Waals crystals~\cite{Novoselov16}, or by subjecting them to external stimuli such as pressure \cite{Badding98}, strain \cite{Schlom07, Zubko11}, or external static fields. More recently, the dynamical manipulation of material properties by optical fields has gained much attention \cite{Mankowsky16, Nicoletti16}. 
Strong laser pulses have been employed to induce metal-insulator transitions~\cite{Rini07, Liu12}, synthetic magnetic fields~\cite{Nova17} and to melt striped phases \cite{Fausti11,Forst14,Cremin19} or charge density waves~\cite{Forst14b, Mankowsky17}. Strong excitation of specific phonon modes can even induce transient superconducting-like phases \cite{Hu14, Mankowsky14, Mitrano16,Cremin19,Budden2020}. 

The coupling to shaped quantum vacua of cavities has been used to change molecular properties \cite{Thomas16,Thomas19, Ebbesen16, Feist18, Ruggenthaler18}, reduce quasiparticle lifetimes in 2D electron gases~\cite{Paravicini19} or drastically enhance the critical temperature of superconductors \cite{Thomas19b}. A growing number of theoretical works investigates this situation~\cite{Curtis18, Sentef18, Kiffner18, Mazza19, Allocca19, Schlawin19, Kiffner19}, where the competition between strong cavity and electron interactions has the potential to manifest in fascinating new physics.
For example, in \cite{Schlawin19} some of us showed how the substantial subwavelength confinement of the cavity field in nanoplasmonic terahertz cavities \cite{Scalari12, Smolka14, Zhang16} can give rise to cavity-mediated electron interactions. A significant drawback of this proposal is the lack of external dynamical experimental controllability of the properties of the interaction potential. 

In this Letter, we combine adjustable external laser driving with the terahertz cavity's shaped vacuum fluctuations to overcome this limitation and create a novel type of controllable long-range electron interactions in 2D materials. 
Importantly, these interactions are manipulated by the parameters of the driving laser: 
their strength is proportional to the laser intensity $I_{\text{d}}$ and inversely proportional to the detuning $\delta_c = \omega_c - \omega_L$ ($\omega_c$ is the cavity frequency and $\omega_L$ is the driving frequency), which allows us to tune them to be attractive or repulsive by changing $\delta_c$. 
Intuitively they can be thought of as the result of inelastic scattering processes in which the intermediate state is dressed by the cavity. 
Strong coherent driving is essential to ensure that the two-photon diamagnetic scattering from which the interactions arise dominates over other (one-photon paramagnetic) processes, while the strong coupling to the cavity field is needed to enhance these interactions but not laser-induced heating processes.
Another advantage of interactions arising from two-photon diamagnetic processes is that they can be induced in a wide range of materials. 
Here we consider a 2D electron gas as a prototypical system. 
We use parameters consistent with semiconductor quantum wells, gated bilayer graphene (BLG) \cite{Cao2018} and the interface between lanthanum aluminate and strontium titanate (LAO/STO) \cite{Caviglia2008} in the normal phase. The latter two are tunable 2D electron gases that can be adjusted arbitrarily close to a spontaneous superconducting instability and exhibit correlated electron phenomena, potentially making the physics even richer. 
Using these examples, we show that the interactions are long-ranged and not effectively screened by the electrons. 
In the attractive regime, we find that they induce an instability in the Cooper channel at experimentally observable temperatures.
 \begin{figure}[t]
     \centering
     \includegraphics[width=.9\linewidth]{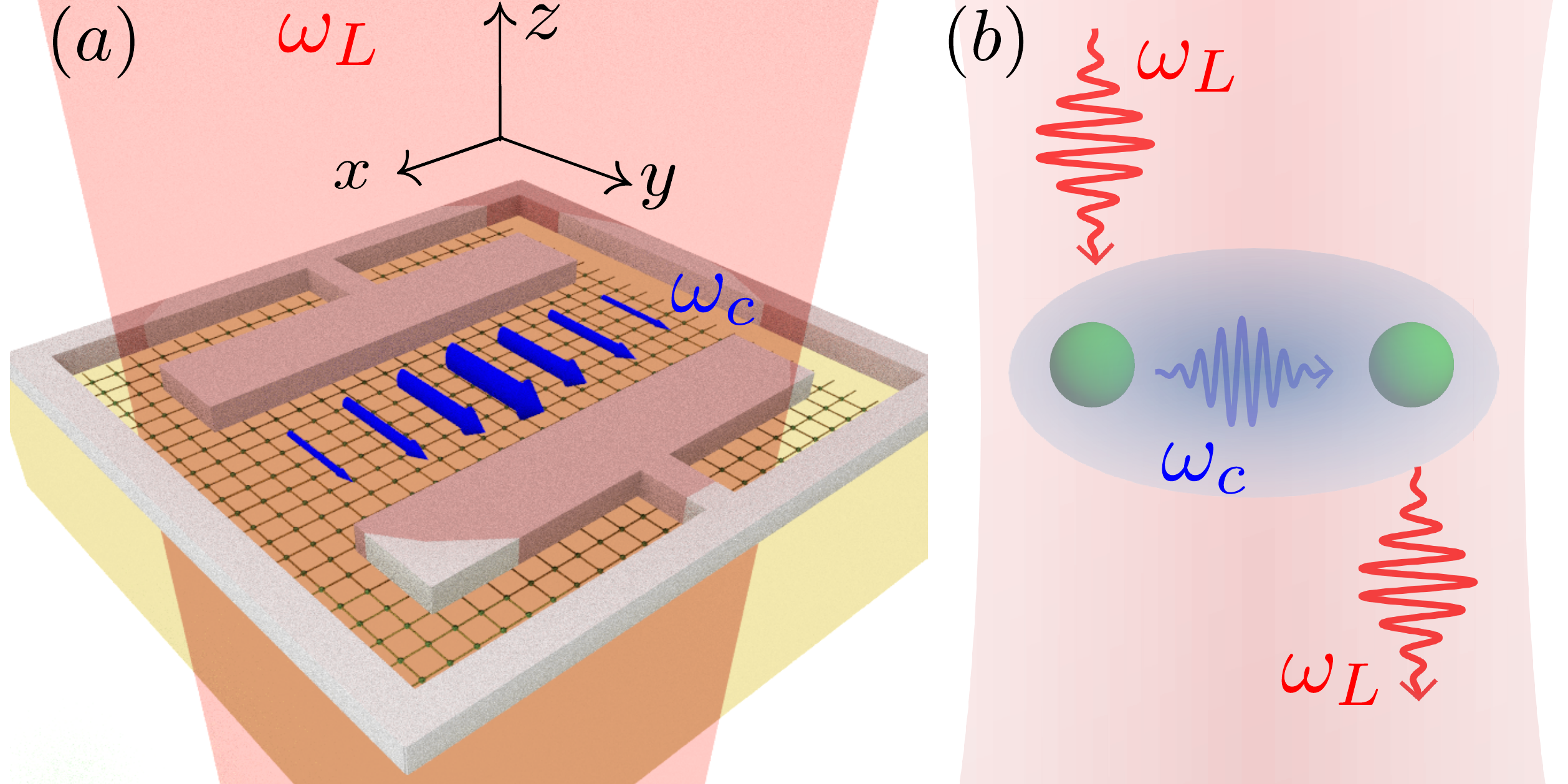}
     \caption{ (a) Setup: 
     a 2D material (indicated by the square lattice) on a substrate (yellow) is coupled to a complementary split-ring cavity (gray) \cite{Maissen14}, whose electric field is represented by the blue arrows with thickness proportional to the field strength. The coupled system is driven by a laser field (schematically shown by the red shading) with wavevector ${\bf q}_L \parallel z$-axis and frequency $\omega_L$ detuned from the cavity frequency $\omega_c$.
     (b) The mechanism of inducing interactions between two electrons (green spheres). Laser(cavity) photons are represented by red(blue) wavy arrows and the green balls are electrons. The red(blue) shading symbolizes the laser(cavity) field.
     }
     \label{fig:setup}
 \end{figure} 
 
\paragraph*{Setup.}
We consider the setup shown in Fig.~\ref{fig:setup}(a). 
A 2D electron system lies in the $xy$-plane inside a substrate material with relative permittivity $\epsilon_r$. It is coupled to the evanescent field of a complementary split-ring cavity which sits on top of the substrate material, as e.g.\ in Ref.~\cite{Maissen14}. This type of cavity is well described by a single-mode light field \cite{footnoteCav} and has been reported to exhibit strong vacuum fluctuations~\cite{Keller17} due to its small cavity mode volume, 
$\mathcal{V}_c = \Lambda \lambda^3$, 
where $\lambda$ is the wavelength of the light mode in the substrate material 
and $\Lambda$ is the mode volume compression factor. Values of $\Lambda=2.5/8\times 10^{-5}$ or even smaller have been reached in experiments and simulations \cite{Maissen14,Kim18}. 
It has also been demonstrated that the cavity is highly reflective only for a narrow bandwidth of frequencies near its resonance, $\omega_c$, and fairly transparent otherwise. 
We note that due to the small sizes of available bilayer graphene samples, LAO/STO may be more straightforwardly integrated into a cavity that extends several hundred microns in size. Furthermore, because LAO/STO structures are insulating at frequencies above the plasma frequency, dissipation may be managed better in these systems than in the graphene structures. 

For simplicity, we thus model the cavity as supporting a single mode and 
perfectly transparent to all frequencies away from $\omega_c$. 
The cavity field is described by a vector potential 
$
{\bf A}_c ({\bf r}) = 2\mathbf{e}_y \sqrt{\hbar/\mathcal{V}_c \epsilon_0 \epsilon_r \omega_c} \cos(q_0 x) \left( b + b^\dagger \right) \text{, }
$
where $\mathbf{e}_y$ is the unit vector in $y$-direction, $q_0 = \omega_c \sqrt{\epsilon_r}/ c$ with $c$ the speed of light in vacuum, $\epsilon_0$ the vacuum permittivity and $b$ the bosonic annihilation operator for the cavity photon.
The cavity-matter system is driven by a strong laser that we describe by an oscillating classical field with vector potential ${\bf A}_d (t) \propto \mathbf{e}_y\sqrt{I_{\text{d}}}\sin(|{{\bf q}}_L| z - \omega_L t)$, where ${\bf q}_L$ is the laser photon wavevector and $t$ is the time. The laser frequency $\omega_L = \omega_c - \delta_c$ lies outside of the cavity reflectivity window, such that the driving field interacts directly with the electronic system.

The coupled cavity-electron Hamiltonian in the Coulomb gauge reads \cite{SuppMat},
\begin{equation}
H = \sum\limits_{j}\frac{(\bm{p}_j+ e{\bf A}_{\text{tot}}(\bm{r}_j,t))^2}{2m} + V_{bg}(\bm{r}_j) + H_{Coul} + H_{cav}
\label{eqn:H full}
\end{equation}
with ${\bf A}_{\text{tot}}(\bm{r}_j,t) = {\bf{A}}_c(\bm{r}_j) + {\bf{A}}_d(\bm{r}_j,t)$. Here $\bm{p}_j$ and $\bm{r}_j$ are the momentum and position operators of electron $j$ with bare mass $m$, $V_{bg}$ is the static potential due to the positive background, $H_{Coul}$ is the Coulomb interaction between the electrons and $H_{cav} = \hbar \omega_c b^{\dagger} b$ is the unperturbed cavity Hamiltonian. 
We assume that the single-electron physics arising from the first part of the Hamiltonian [\eqr{eqn:H full}] is described by a single band and higher bands are far detuned from the cavity frequencies considered here ($\lesssim$ 10 THz) such that dynamical Stark effects are negligible. 
The first term in \eqr{eqn:H full} also yields one-photon paramagnetic ($ \bm{p} \cdot {\bf A}$) and two-photon diamagnetic ($ {\bf A}^2$) interactions between the transverse light fields and the electrons. 
In \eqr{eqn:H full} we neglected the explicit description of environmental EM field modes, though we checked that for the parameters we consider their main influence---inelastic scattering of laser photons into the environment---does not heat the electrons significantly. This is because the engineered interactions scale as $1/(\Lambda\delta_c)$ [see \eqr{eqn:BareIntMatrixElement}] whereas the inelastic scattering rate scales as $1/\omega_L \ll 1/(\Lambda\delta_c)$ \cite{SuppMat} and thus affects the electron dynamics only very weakly. 

\paragraph*{Rotating-frame transformation.}
Within the single electronic band, the one-photon processes which normally dominate the optical response of materials are far off-resonant. 
However, the two-photon diamagnetic interactions include near-resonant processes [schematically shown in \fir{fig:setup}(b)] in which laser photons are scattered into the cavity. When the driving is strong \cite{SuppMat}, these processes become the principal coupling mechanism of the electrons to the cavity field fluctuations and dominate the dynamics.
This becomes clear when we move into a frame co-rotating with the laser frequency $\omega_L$ using the unitary transformation 
$U (t) = \exp ( - i \omega_L t b^\dagger b)$, see \fir{fig:screening}(a).
This transformation does not affect the electronic Hamiltonian but changes $\omega_c$ in $H_{cav}$ to $ \delta_c$ ($|\delta_c |\ll \omega_c$) and removes the explicit time dependence of the near-resonant two-photon diamagnetic interactions. In this frame, all the one-photon interaction terms oscillate rapidly at $\omega_L$, thus their effect on the electrons is proportional to $\sim 1/\omega_L$. 

In contrast, the effect of two-photon diamagnetic interactions (apart from an irrelevant energy shift) is proportional to $\sim 1/\delta_c \gg 1/\omega_L$. 
In this rotating frame, the near-resonant part of the two-photon diamagnetic interactions is given by
\begin{equation}
H_{\text{light-matter}} = \sum\limits_{{\bf q}=\pm q_0 \mathbf{e}_x } \frac{ g_0 }{\sqrt{\mathcal{S}}} \varrho_{\bf q} (b +b^\dagger) \text{ ,}
\label{eq.H_int}
\end{equation}
where $\mathcal{S}$ is the cavity area in the $xy$-plane and $\varrho_{\bf q} = \sum_{{\bf k},\sigma} c^\dagger_{{\bf k + q},\sigma} c_{\ksig}$ is the electron area density in $k$-space. $c_{\ksig}$ annihilates an electron with quasi-momentum ${\bf k}$ and spin $\sigma$. 
The interaction vertex reads
$
g_0 = \alpha \sqrt{I_{ \text{d} } \hbar^3 \omega_c^2 \mathcal{S} / (2\pi m^2 \Lambda \omega_L^2 c^2)} \text{,}
$
where $\alpha$ is the fine-structure constant. 

\paragraph*{Cavity-mediated interactions.}
The light-matter coupling given by Eq.~(\ref{eq.H_int}) yields effective electron interactions through the process depicted in \fir{fig:setup}(b):
a laser photon scatters virtually off an electron into the cavity, where it gets rescattered by a second electron back into the laser beam. 
The fact that the two-photon diamagnetic interactions couple to the electron density [see \eqr{eq.H_int}] allows us to apply some of the intuition from other boson-exchange interactions \cite{Carbotte90}. 
Specifically, it means the exchange of virtual cavity photons (with effective frequencies $\delta_{c}$ in the rotating frame) mediates density-density interactions between the electrons. The interactions will be attractive(repulsive) when $\delta_{c} > 0(<0)$, $i.e.$ the driving laser is red(blue)-detuned from the cavity.

\begin{figure}[t!]
     \includegraphics[width=.93\linewidth,angle=0,origin=c]{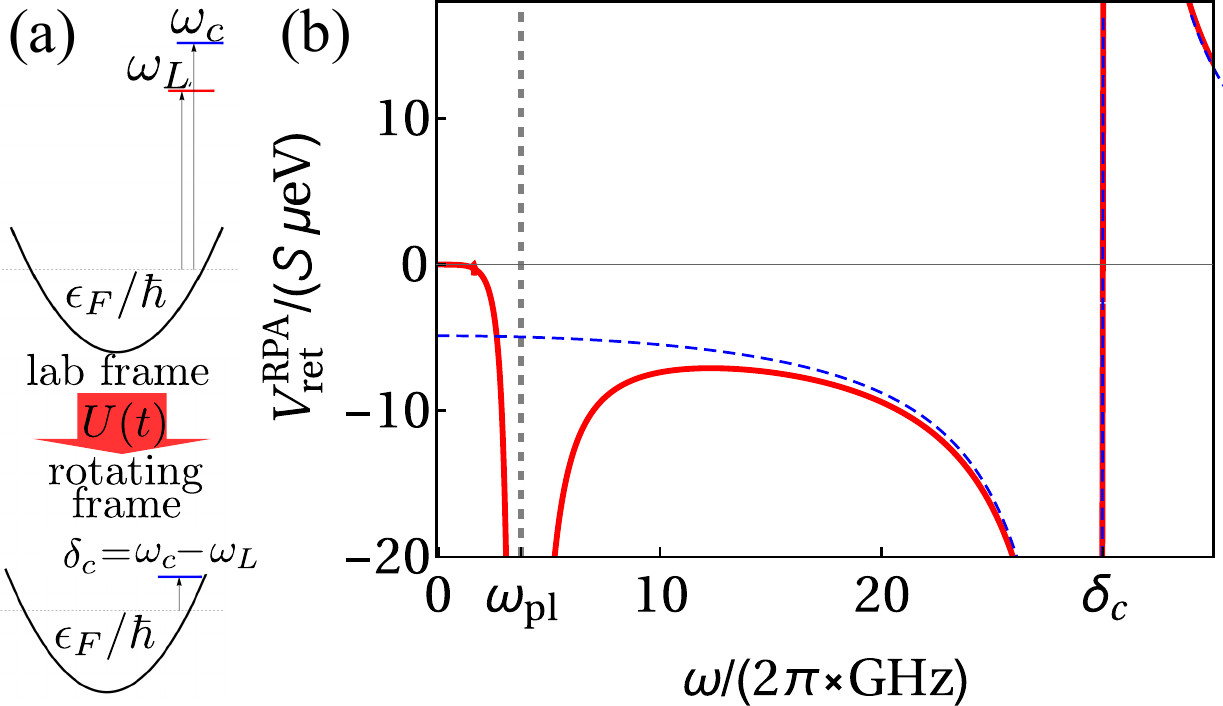}
     \caption{(a) Relevant energy levels for the two-photon diamagnetic interactions involving the driving (red) and cavity (blue) fields in the lab frame (top) and in a rotating frame (bottom). The electron dispersion is plotted in black. $\epsilon_F$ is the Fermi level. 
     (b) The real part of the screened interaction potential $V^{\text{RPA}}_{\text{ret}}/\mathcal{S}$ is shown in red. 
     The real part of $V^{\text{bare}}_{\text{ret}}/\mathcal{S}$ (blue dashed line) is shown for comparison. The plasmon resonance $\omega_{pl}(q_0)$ is indicated by the grey vertical dashed line. 
     The small irregularity below $\omega_{pl}$ comes from the discontinuity of the 2D electron polarizability given in Ref.~\cite{Stern67}. 
     The cavity parameters used are $\omega_c = 2 \pi \times 0.3$ THz, $\delta_c = 0.1 \omega_c$, quality factor $Q=500$ and $\Lambda = 2.5/8\times 10^{-5}$. We set $\mathcal{S} \approx 2.1\times 10^{-5} \lambda^2$, where the proportionality constant is obtained from Ref.~\cite{Maissen14}. 
     The driving intensity is $I_{\text{d}} = 3\text{MWcm}^{-2}$. 
     We use an electron density $n_e = 10^{12}$cm$^{-2}$ and $m^{*}=2m$ consistent with various 2D materials such at (gated) LAO/STO and monolayer TMDs \cite{McCollam14,Thiel06,2DMaterial_Rasmussen15}. The plasmon damping rate is set to $10$\% of plasmon frequency \cite{Gusikhin2015}. 
     We consider a $\text{SrTiO}_3$ substrate with a large $\epsilon_r \sim 10^4$ at low temperatures \cite{Sakudo71}.}
     \label{fig:screening}
 \end{figure}

The bare retarded interaction created by the exchange of cavity photons reads $V^{\text{bare}}_{\text{ret}}({{\bf q}},\omega+ i \eta) = |g_0|^2 D^{(0)}({{\bf q}},\omega+ i \eta)$, where $\eta=0^{+}$ and $D^{(0)}$ is the bare cavity photon Green's function in the rotating frame. 
In the static limit $\omega = 0$, we recover the effective interaction Hamiltonian obtained using a Schrieffer-Wolff transformation, 
$H_{\text{int}} = \frac{1}{2\mathcal{S}}\sum_
{{\bf{q}}} 
V^{(0)}({{\bf q}}) \varrho_{{\bf q}} \varrho_{-{\bf q}}$, 
where the expression for the bare interaction potential, 
\begin{equation}
\frac{V^{(0)}({\bf q})}{\mathcal{S}} = \frac{\alpha^2 \hbar^2 }{\pi c^2 m^2 } \frac{\omega_c^2}{\omega_L^2} \frac{I_{\text{d}}}{  \Lambda \delta_c} \delta_{\pm {\bf q},q_0 \mathbf{e}_x} \text{,}
\label{eqn:BareIntMatrixElement}
\end{equation}
confirms our intuition that the interaction strength is proportional to the driving intensity and can be tuned to be attractive or repulsive by the detuning. 
We note that this potential only depends on the momentum transferred between the two electrons and does not depend on the electron dispersion. 
Moreover, $V^{(0)}({\bf q})$ is very localized in momentum space because the photon momenta are much smaller than electronic ones. Hence, this electron interaction is long-ranged in real space. 
In the following section, we will investigate how it is screened by the two-dimensional electron gas.

\paragraph*{Screening.} 

Within the random phase approximation (RPA), the screened interaction potential in imaginary time formalism reads \cite{BruusFlensberg04},
\begin{equation}
V^{\text{RPA}} (\tilde{q} ) =  \frac{ \left\vert g^{\text{RPA}}(\tilde{q}) \right\vert^2 D^{(0)} (\tilde{q}) }{ 1 - \left\vert g_{0} \right\vert^2 \chi^{\text{RPA}} (\tilde{q}) D^{(0)} (\tilde{q}) }, \label{eq.screened}
\end{equation}
where we have used the four-vector notation with $\tilde{q}\equiv \lbrace {\bf q}, i \nu_n\rbrace$ at momentum $\bf q$ and the bosonic Matsubara frequency, $i\nu_n=2\pi n k_\text{B} T/\hbar$, $n \in \mathds{Z}$. $k_\text{B}$ is the Boltzmann constant and $T$ is the temperature. $\chi^{\text{RPA}} $ is the screened polarizability of the two-dimensional electron gas, and $ g^{\text{RPA}} $ is the screened vertex of Eq.~(\ref{eq.H_int}). We obtain $\chi^{\text{RPA}}$ and $ g^{\text{RPA}} $ for a parabolic band with an effective mass $m^{*}$ using results calculated in~\cite{Stern67} and we present their expressions in the SM. Since $q_0 \ll k_f$, we are in the long-wavelength limit where only electrons near the Fermi surface contribute to screening. Thus, the results in \cite{Stern67} are valid in this limit for other single band dispersions \cite{Hwang07}. The retarded interaction is obtained from \eqr{eq.screened} (a function in imaginary frequency) by analytic continuation to real frequencies \cite{Mahan}:  $V^{\text{RPA}}_{\text{ret}}(\omega) = V^{\text{RPA}}(\pm q_0 \mathbf{e}_x, i \nu_n \rightarrow \omega + i \eta) $.

We show an example of the real part of the screened retarded interaction potential as a function of frequency in \fir{fig:screening}(b). Here we also included finite Drude-like electron \cite{Fox2001,kittel1987quantum} and cavity damping rates (see SM for details). 
Evidently, contributions to the interaction at small frequencies are strongly screened. 
The precise choice of the damping rates alters only the size of the dips at $\omega=\omega_{\text{pl}}$ and $\delta_c$ but not the interaction potential in the broad region of frequency in between, thus it has little effect on the physics.
In the present case where $\delta_c > 0$ the attractive cavity-mediated interaction softens the plasmons and the cavity frequency is weakly blue-shifted. This effect only becomes appreciable at very large laser intensities, and cannot be seen in \fir{fig:screening}(b).
Crucially, however, for frequencies between the plasmon resonance and the detuning $\delta_{c}$, 
the interaction remains attractive (repulsive for $\delta_{c} < 0$) and is hardly affected by screening. 
We can understand this behaviour by investigating the expression for the screened interaction in more detail.

\begin{figure}[t]
\centering
\includegraphics[width=\linewidth]{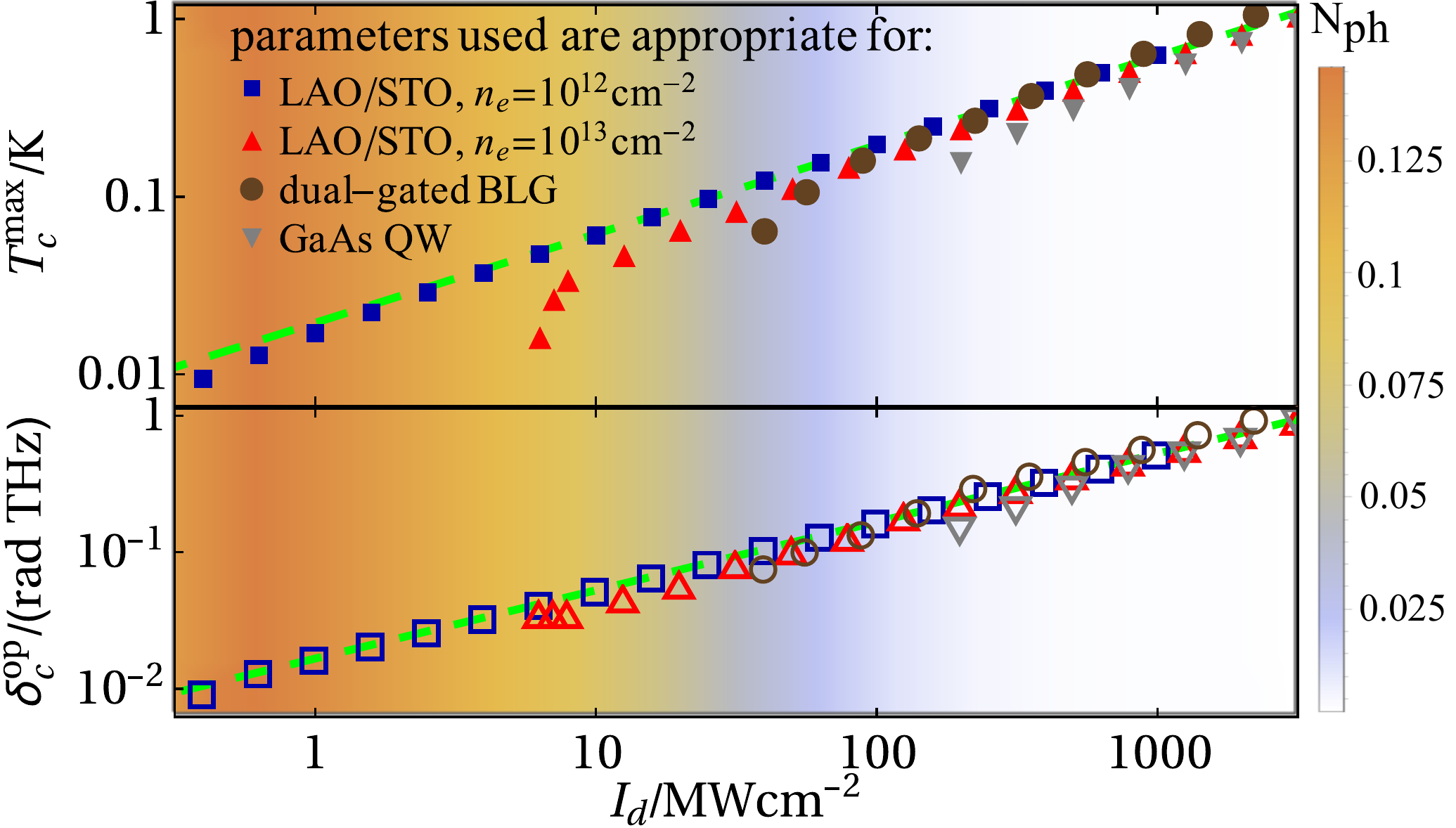}
\caption{Maximum $T_c$'s (w.r.t.\ driving and cavity parameters) for the Cooper channel instability (top panel) and the optimal detunings, $\delta_c^{\text{op}}$, (bottom panel) are plotted for a range of driving intensities.
We calculate them for four sets of electron parameters. All of them follow the dashed green line (same in both panels) representing \eqr{eqn: proportionality} at large enough driving intensities. 
The colour map shows the cavity photon number $\text{N}_{\text{ph}}$ for different $I_{\text{d}}$ with the optimal driving and cavity parameters at $T_c^{\text{max}}$ using electron and substrate parameters appropriate for gated LAO/STO [same as those used in \fir{fig:screening}(b)]. We used $n_e= 3\times10^{11}$, $5\times10^{10}\text{cm}^{-2}$, $\epsilon_r= 10$, $13$ for dual-gated BLG and GaAs based quantum well (QW) respectively \cite{Yan2011,Robertson04,Zhang16}; $m^{*} = 0.07m$ for GaAs QW \cite{Zhang16}. AB-stacked BLG dispersion is calculated as in Ref.~\cite{McCann2007} with a gate-induced band gap of $50$meV. 
We used $ \omega_c = 10\text{ } \delta_c^{\text{op}} $. 
}
\label{fig: Max Tc plot}
\end{figure}

Since we are in the long-wavelength limit, we have $\chi^{\text{RPA}} \propto q_0^2$. 
Physically this means the electrons are too slow to screen the fast oscillating electromagnetic fields and hence 
the cavity-mediated interactions can only be screened weakly. 
Corrections to the bare interaction become appreciable when $\omega$ approaches the plasmon resonance of the system. In a two-dimensional electron gas, the plasmon dispersion is given by $\omega_{pl}^2 (q_0) = e^2 k_f v_f q_0/ (4 \pi \hbar \epsilon_r \epsilon_0)$ \cite{AndoRevModPhys82}, where $v_f$ is the Fermi velocity. This yields plasmon frequencies in the range $\omega_{pl}/2\pi \sim 4$~GHz [for the parameters of \fir{fig:screening}(b)]. Hence, it is only relevant at temperatures below $T \simeq 30$mK, as we will see in \fir{fig: Max Tc plot}. At higher temperatures, the Matsubara frequencies will not cover this low-frequency region and the cavity-mediated interactions cannot be screened effectively by the electron gas. They remain long-ranged interactions that can be imposed externally on the two-dimensional electronic system.

\paragraph*{Cooper channel instability.}
To gauge the strength of the cavity-mediated interactions that could be achieved experimentally, we consider the case of attractive interactions and study the Cooper instability they induce. This pairing instability occurs when a pair of electrons with opposite momenta and frequencies attract and repeatedly scatter off one another to cause a divergence of the pair scattering vertex $\Gamma(\tilde{k};\tilde{p})$, where $\tilde{k}$ ($\tilde{p}$) is the incoming (outgoing) electron four-momentum \cite{BruusFlensberg04,abrikosov1975methods}. The instability could potentially be detected with techniques such as transport or magnetisation measurements \cite{Reyren07,Paravicini19,McIver2020}.
$\Gamma(\tilde{k};\tilde{p})$ is obtained by solving a Dyson equation \cite{BruusFlensberg04, SuppMat}. 
As the photon momentum is very small compared to electronic momenta, we approximate the exact interaction by a $\delta$-function potential in $k$-space, $V^{\text{RPA}} (\tilde{q} ) \propto \delta ({\bf q})$, which simplifies the internal summations in the pair scattering ladder diagrams, yielding a matrix equation (see SM for more details)
$    \Gamma(i k_n;ip_n) = - V^{\text{RPA}}(i k_n-ip_n) + \mathcal{M}(i k_n;i\nu_n) \Gamma(i\nu_n;ip_n)$, 
where $\mathcal{M}(i k_n;i\nu_n) = - 2 V^{\text{RPA}}(i k_n-i\nu_n) G^{(0)} (i\nu_n) G^{(0)} (- i\nu_n) / (\mathcal{S} \beta)$, and $G^{(0)}$ is the bare electron Green's function and $\beta=1/k_\text{B}T$ is the inverse temperature. 
We then numerically compute the temperature $T_c$, at which the pair scattering vertex diverges using the Dyson equation with the screened interaction potential within RPA for different choices of driving and cavity parameters. 

$T_c$ depends on the driving intensity, the cavity frequency and the detuning. In the following, we vary the detuning to obtain the highest critical temperature for fixed intensities. The results are shown in \fir{fig: Max Tc plot}, where the highest $T_c$, $T_c^\text{max}$ (top panel), and the corresponding detuning $\delta_c^{\text{op}}$ (bottom panel) are plotted against the intensity for four sets of electron and substrate parameters. 
We find that provided $\omega_c \gg \delta_c$ and $\delta_c$ is well above the plasmon resonance $T_c$ depends very weakly on $\omega_c$ once we fix $\delta_c$: a smaller $\omega_c$ raises $T_c$ very slightly. 
We remark that the necessary intensities to induce $T_c \sim 1$~K are experimentally achievable with pulsed THz light sources \cite{Hafez_2016,Yeh07,Hoffmann:09,Liu12,Vicario2013,Kubacka1333}. At these high intensities, laser-induced heating of the substrate material becomes relevant \cite{SuppMat}, such that a pulse driving protocol might become necessary. 
The green dashed line represents a function $\propto \sqrt{I_{\text{d}}}$ and it fits the $T_c^{\text{max}}$ and the $\delta_c^{\text{op}}$ points very well at high driving intensities, 
\begin{equation}
T_c^{\text{max}}/ \text{K} \approx \delta_c^{\text{op}} \text{ ps}/\text{rad} \approx 10^{-4.5} \sqrt{I_{\text{d}}/ (\Lambda \text{ MWcm}^{-2})} \text{.}
\label{eqn: proportionality}
\end{equation}
The numerical constants depend only on fundamental constants. The deviations of our numerical results in \fir{fig: Max Tc plot} from \eqr{eqn: proportionality} at smaller intensities are caused by the screening due to plasmons which reduces $T_c^{\text{max}}$. Consequently, the onset of this deviation is determined by the plasmon frequency and hence the substrate dielectric constant, the 2D electron density and the Fermi velocity. The materials with fewer data points in \fir{fig: Max Tc plot} have higher plasmon resonances, so the screening is stronger for them, causing $T_c^{\text{max}}$ to drop quickly to 0 at higher intensities than materials with lower plasmon resonances.

\fir{fig: Max Tc plot} shows that in order to achieve the highest critical temperature at a given driving intensity, one must choose an appropriate detuning (and hence cavity and driving frequencies). 
We understand this by applying an approximation to the Dyson equation to estimate the Cooper instability temperature. The approximation, which underestimates $T_c$ by only a few percent, amounts to retaining only the two fermionic Matsubara frequencies of the lowest magnitude in the Dyson equation
\cite{BruusFlensberg04}. 
This allows us to extract the critical temperature analytically, 
$k_B T_c \approx -2\pi^{-2} V^{RPA}(q_0 \mathbf{e}_x, 2\text{i}\pi k_B T_c/\hbar) /\mathcal{S}  $. 
Above the plasmon resonance, the interaction is effectively unscreened so for a perfect cavity $V^{RPA} \propto I_{\text{d}} \delta_c/((2\pi k_B T_c/\hbar)^2+\delta_c^2)$ \cite{SuppMat}. For all values of $T_c$, this is maximised when we choose $\delta_c^{\text{op}} = 2\pi k_B T_c/\hbar$. 
We substitute the two relations above back into the approximate equation for the critical temperature to obtain $T_c^{\text{max}} \propto \sqrt{I_{\text{d}}}$ consistent with \fir{fig: Max Tc plot} and \eqr{eqn: proportionality}. 
We note that a finite cavity decay rate reduces $\delta_c^{\text{op}}$ only slightly.

We furthermore verified that the cavity photon population $\text{N}_{\text{ph}}$ remains low at $T_c$ (see colorbar in \fir{fig: Max Tc plot}) when we use the optimal cavity and driving parameters \cite{SuppMat}. This justifies neglecting two-photon diamagnetic interactions involving only the cavity modes (${\bf A}_c^2$ terms). The decrease of $\text{N}_{\text{ph}}$ with higher driving intensity is due to the blue-shift of the cavity resonance (see the above discussion on screening) resulting in a lower thermal cavity population. 
Finally, we note that keeping $\delta_c$ well above the cavity linewidth and the plasmon frequency suppresses the effect of cavity-induced heating or cooling \cite{Piazza14}.

\paragraph*{Conclusions.}
We showed that the combination of external driving and strong coupling to a cavity induces controllable long-range electron interactions in single-band 2D materials. Provided that the driving is detuned from the cavity further than the plasmon frequencies at the relevant wavevectors, the interactions cannot be effectively screened. The interactions are generated by two-photon diamagnetic processes. Therefore, they are largely independent of the electron band dispersion and can be induced on top of existing interactions. 
These induced interactions could be used to control or enhance existing instabilities or give rise to different quantum states through competition with other short-range interactions \cite{Landig2016,Camacho-Guardian17,Fan18,Schlawin19atom}.
To date, engineered long-range quantum fluctuation-mediated interactions have mainly been studied in the context of ultracold atoms \cite{Jaksch01,Munstermann00,Maschler05,Baumann2010,Mottl12,Landig2016,Zeytinoglu16,Mivehvar17,Kroeze18,Vaidya18,Norcia259,Guo19,Mivehvar19}. 
Our proposal potentially opens the possibility to explore a vast range of novel physics related to unscreened controllable long-range interactions in condensed matter setups. 

\paragraph*{Acknowledgements}
We would like to thank A.~Imamo\u{g}lu, J.~Faist, G.~Mazza and C.~S\'anchez Mu\~noz for useful discussions.
This work has been supported by the European Research Council under the European Union's Seventh Framework Programme (FP7/2007-2013)/ERC Grant Agreement No.\ 319286 Q-MAC and by EPSRC grant No.\ EP/P009565/1.
We acknowledge support from the Deutsche Forschungsgemeinschaft (DFG) via the Cluster of Excellence “CUI: Advanced Imaging of Matter” (EXC
2056 - Project ID 390715994) and the Collaborative Research Center SFB 925.


\begin{thebibliography}{93}
\bibitem [{foo()}]{footnoteCav}%
  \BibitemOpen
  \href@noop {} {}\bibinfo {note} {The next higher mode has a frequency of
  roughly $4\omega_c$ and is more lossy \cite{Maissen14}.}\BibitemShut {Stop}%
\end{thebibliography}

\putbib
\end{bibunit}
\end{document}